*Engineering Photonic Environments for Two-dimensional Materials*

*Xuezhi Ma[1], Nathan Youngblood[2], Xiaoze Liu[3], Yan Cheng[4], Preston Cunha[1], Kaushik Kudtarkar[1], Xiaomu Wang[4], and Shoufeng Lan[1*]*

**Addresses:**

[1] *Department of Mechanical Engineering, Texas A&M University, College Station, TX 77840, USA*

[2] *Department of Electrical and Computer Engineering, University of Pittsburgh, Pittsburgh, PA 15261, USA*

[3] *School of Physics and Technology, Wuhan University, Wuhan, Hubei, 430072, China*

[4] *School of Electronic Science and Engineering, Nanjing University, Nanjing, Jiangsu, 210023, China.*

Corresponding author information: shoufeng@tamu.edu

**Abstract:**

A fascinating photonic platform with a small device scale, fast operating speed, as well as low energy consumption is two-dimensional (2D) materials, thanks to their in-plane crystalline structures and out-of-plane quantum confinement. The key to further advancement in this research field is the ability to modify the optical properties of the 2D materials. The modifications typically come from the materials themselves, for example, altering their chemical compositions. This article reviews a comparably less explored but promising means, through engineering the photonic surroundings. Rather than modifying materials themselves, this means manipulates the dielectric and metallic environments, both uniform and nanostructured, that directly interact with the materials. For 2D materials that are only one or a few atoms thick, the interaction with the environment can be remarkably efficient. This review summarizes the three degrees of freedom of this interaction: weak coupling, strong coupling, and multi-functionality. Also, it reviews a relatively timing concept of engineering that directly applied to the 2D materials by patterning. Benefiting from the burgeoning development of nanophotonics, the engineering of photonic environments provides a versatile and creative methodology of reshaping light-matter interaction in 2D materials.

1. Introduction

It has been golden years of expansion to the family of two-dimensional (2D) materials since the dormancy of atomically thin graphene more than a decade ago [1]. The tendency of increasing in terms of the number of family members of the 2D materials and research interests, in general, will continue [2, 3]. The reason that researchers are passionate about studying these 2D materials is their unique combination of features having both strong in-plane bonds and relatively weak out-of-plane van der Waals forces. Those features have made it possible to exfoliate layered 2D materials down to one or a few atoms in thickness. While initially studied for their superior electronic properties [1], the photonic counterparts are becoming increasingly important. Many of those photonic properties have even been icons of the corresponding 2D materials. For example, the optical absorption of single-layer graphene in a broad wavelength range is 2.3%, defined solely by the fine structure constant, regardless of the material parameters [4]. A critical factor for the development of the research field in 2D material-based photonics is the ability to modify and control the optical responses beyond those initially offered by the 2D materials, for instance, to increase the optical absorption of single-layer graphene to near unity.

A promising means by which one can modify the optical properties of 2D materials are engineering the photonic environment. This method is benefiting greatly from the advancement of nanofabrication. Early research in this field treats the photonic environment as a necessary holder for measurements and integration. For example, metal pads are used as electrodes for conducting electrical signals, dielectric layers can be spacers for applying electrical potentials, and wafers function as substrates for holding the 2D materials. With the advancement of a distinct research field, micro- and nano-photonics, the photonic environments are playing more and more crucial roles in 2D material-based photonics [5]. In general, we can sort the roles they play with 2D materials into three categories: weak coupling (Purcell effect), strong coupling (polaritonics), and multi-functionalities. Enhancement refers to the increase in the optical density of states or strengthening in light-matter interactions [6-9]. According to the Purcell effect, the enhancement factor is proportional to the ratio of the quality factor and mode volume [10]. Therefore, shrinking mode volumes in real space with the confinement of optical modes can also be a way to enhance light-matter interactions. For multi-functionalities, the photonic environments can introduce functionalities that may not originally belong to 2D materials, such as the valley Hall effect for separating the valley polarization [11, 12]. The role photonic environments play is also closely associated with their physical structures.

A waveguide typically contains a core with a high refractive index, which is surrounded by a low refractive index [13]. The index contrast forces the light to propagate within the core area, governed by the total internal reflection. Depending on the geometry, the propagating light can cover a wide range of wavelengths in the electromagnetic spectrum. When the two terminals connect, the waveguide can transform into a loop or so-called resonator, which supports optical standing modes [14]. Depending on the size of the resonator, the standing modes will resonant at different specific wavelengths. More importantly, the quality factor indicates the ability of the resonator to store optical energy. This factor can reach a value of a million or more, which is critical for phenomena that require a high energy density. Wavelength selectivity and the ability to store optical energy are also achievable in periodic structures that form a photonic band structure under Bloch's theorem [15]. One-dimensional periodic structures can create a distributed Bragg reflector (DBR) that functions as a mirror [16]. Two-dimensional periodic structures are so-called photonic crystals since they support a photonic band structure, analogous to electronic band structures in natural crystals [17]. The periodicity of the photonic crystals is typically comparable to the wavelength of interest [18], in relatively new metamaterials, on the other hand, the periodicity can be much smaller than the wavelength. More importantly, the methodology for obtaining material properties by designing the metamaterials' structures and geometries rather than their chemical compounds of the building blocks, holds enormous possibilities to modify materials at will [19]. For 2D materials, the building blocks can also independently create an electronic metamaterial by patterning [20].

In this article, we review the recent advancement of this research frontier in 2D material-based nanophotonics. Our focus is on the engineering of the photonic environment that surrounds the 2D materials in four aspects: Purcell effect in the weak coupling regime, polaritonics in the strong coupling regime, multi-functionalities using metamaterials, and the direct patterning of 2D materials. We will only briefly present the optical properties and functionalities of 2D materials.

## 2. Enhancement through photonic integration

Integration of 2D materials with photonic circuits provides numerous advantages such as enhanced quantum yield [21-24], improved collection efficiency of emitted photons [25-28], easy routing and manipulation of optical signals [29, 30], enhanced light-matter interaction [9, 31-33], and the formation of exciton-polaritons (EPs) states [34]. Due to out-of-plane van der Waals (vdW) interactions, 2D materials do not require lattice matching or buffer layers and therefore can be placed directly on arbitrary substrates owing to those advanced 2D materials transfer methods [35, 36]. An additional benefit of photonic integration is the decoupling of the interaction length from the material's thickness via coupling to the evanescent fields of a photonic waveguide or resonator [37]. The scenario of the integration of 2D materials with photonic environments can be treated as the coupling of light and 2D materials in cavity systems [34, 38, 39], where the photon modes are in resonance with the excited states of 2D materials. While the excited states refer to 2D excitons in transition metal dichalcogenides (TMDs) for most cases, various cavities have been realized to enhance the coupling with sophisticated nanotechnologies [34, 38, 39]. Depending on the coupling strength, this coupling can be defined as weak coupling or strong coupling regime [40-43]. In this chapter, we will focus on the week coupling regime, in which the interaction between the excitons and photons is irreversible in the cavities [40, 41]. Through this coupling, the energy can only transfer from the excitons to the photons [40, 41]. We will discuss the strong coupling regime in the next chapter.

### 2.1 Enhancing quantum yield:

Compared with other bulk semiconductors, 2D materials, such as transition metal dichalcogenides (TMDs), typically suffer from low quantum yield [44]—especially at room temperature. We can define the quantum yield as:

$$\eta = \gamma^{rad}/(\gamma^{rad} + \gamma^{loss}) \qquad (1)$$

where $\gamma^{rad}$ and $\gamma^{loss}$ are the radiative and nonradiative recombination rates of excitons in the 2D material, respectively. This low yield ($\eta \sim 0.1\%$ to $10\%$) stems from the atomically thin nature of these materials, which causes defects [45], surface interactions [46], and other excitons [47] to dominate nonradiative recombination rates in these materials (i.e. $\gamma^{loss} \gg \gamma^{rad}$). Chemical doping is one method which has proven highly effective for enhancing quantum yield by passivating surface defects and significantly reducing the nonradiative recombination rate in MoS₂ [48]. However, in addition to decreasing $\gamma^{loss}$, it is also possible to increase $\gamma^{rad}$ by modifying the PDOS of the environment surrounding the 2D material such that $\gamma^{rad} > \gamma^{loss}$. This is accomplished by integrating the 2D material into a resonant optical cavity, such as a photonic crystal, micro-ring resonator, or Fabry-Perot resonator, which enhance the spontaneous decay rate at certain frequencies while inhibiting it at other frequencies. This enhancement is known as the Purcell effect, which can be defined as follows in the case of weak coupling to the cavity [49]:

$$F_p \equiv \frac{\gamma_{ex}}{\gamma_{ex}^0} = \frac{3}{4\pi^2}\left(\frac{\lambda}{n}\right)^3 \frac{Q}{V_{eff}} \quad (2)$$

where $F_p$ is the Purcell factor, $\gamma_{ex} = \gamma^{rad} + \gamma^{loss}$ is the total exciton decay rate in the 2D material, $\gamma_{ex}^0$ is the radiative decay rate of the exciton in free space, $\lambda$ is the emitted wavelength, $n$ is the refractive index of the environment, $Q$ is the quality factor of the resonator, and $V_{eff}$ is the effective mode volume. It is important to note that since the 2D material is placed in close proximity to the optical cavity, both $\gamma^{rad}$ and $\gamma^{loss}$ will be modified. Therefore, losses from the cavity can undesirably increase $\gamma^{loss}$ like in the case of dipole emission quenching due to metallic ohmic losses in plasmonic resonators [50]. The modified quantum yield of excitonic emission from a 2D material coupled to a nanophotonic cavity can be written [49, 51]:

$$\eta = \eta_0 \frac{F_p^{rad}}{\eta_0 F_p + (1-\eta_0)} \quad (3)$$

where $\eta_0$ is the quantum yield of the isolated excitonic system in free space and $F_p^{rad} = \gamma^{rad}/\gamma_{ex}^0$ is the Purcell enhancement of radiative emission due to the cavity. The above equation is encouraging because it allows one to modify the quantum efficiency of excitonic emission by engineering the nanophotonic cavity's Purcell factor (i.e. maximizing $Q/V_{eff}$). In the next two sections, we discuss various approaches to enhancing the quantum yield in 2D materials using both dielectric and plasmonic resonators and discuss the possibility of combining these resonators with larger photonic integrated circuits.

2.2 Dielectric Structures

Some of the first examples of spontaneous emission enhancement using a nanophotonic cavity were from studies of photoluminescence (PL) in MoS₂ [21, 22, 52]. Prior works identified the low quantum yield (QY) of room temperature PL ($\eta_{PL} \sim 0.01$) stemming from the significant difference in the nonradiative versus radiative lifetimes of the excitons in MoS$_2$ ($1/\gamma^{loss} \sim 100$ ps compared to $1/\gamma^{rad} \sim 10$ ns) [46, 53]. In 2013, Gan et al. demonstrated the ability to control the spontaneous emission rate of MoS$_2$ by placing it on a photonic crystal cavity (PCC), which increased the radiative recombination rate (and therefore, the quantum yield) by a factor of $F_p \sim 70$ [21]. Placing the 2D material on the low mode volume, PCC enhanced not only the Purcell factor but also the out-of-plane emission (and therefore PL collection efficiency) by suppressing in-plane propagating modes which lay within the photonic bandgap of the crystal [54]. This enhancement of directionality is an additional benefit that can be gained from engineering the PDOS of photonic systems [21, 22, 26, 55].

In the context of coherent light-emitting devices (e.g., lasers and single-photon emitters), resonators with a small mode volume and a narrow band of allowed photonic modes are desirable for lowering the lasing threshold. In this context, PCCs [56], micro-disk resonators [57, 58], and vertical Fabry-Perot cavities [59] have all been used to achieve resonant enhancement of stimulated emission through the Purcell effect. An excitonic nanocavity laser based on a monolayer of WSe$_2$ integrated on a PCC was first demonstrated by Wu et al. in 2015 (Figure 2A) [56]. In this work, the PCC was reduced to a thickness of 125 nm, which served to enhance the nanocavity's $Q$, decrease the effective mode volume $V_{eff}$, and enhance the evanescent fields overlap with the WSe$_2$, improving the Purcell factor and reducing the lasing threshold of the device. While room temperature lasing was not demonstrated in this device, other works have shown the feasibility of room temperature lasing in monolayer TMDs materials [57, 59]. This works claimed with a Q factor of 2500 and large F$_p$ of 37, while a micro-disk laser of monolayer WS$_2$ was claimed with a of 2604 (Figure 2C) [58]. In the meantime, another microsphere laser of monolayer MoS$_2$ was also reported with Q in the range of 2600 to 3300 (Figure 2D) [60]. These early examples have shown the high is necessary to reach the lasing regime with the superlinear pump-power dependence and narrowing of lasing linewidths [34, 38]. Later, a nanobeam cavity based on monolayer MoTe$_2$ was reported as the first room-temperature continuous-wave (CW) lasing for 2D TMDs with an ultrahigh of 5603 (Figure 2E) [61]. Other than realizing high- cavity, another effective way to reach the lasing regime is to engineer the QY of 2D excitons so that the requirement for a high-Q cavity could be litigated [62]. As a result, a vertical-cavity surface-emitting laser (VCSEL) of monolayer WS$_2$ was also realized with only a moderate Q of 640 (Figure 2F) [59]. Along with these progresses, the lasing regime through the weak light-matter coupling is still under investigations since some critical information for 2D lasers is still unclear [62, 63]. For example, some of the outstanding questions are how much the linewidth is narrowed and how is the coherence of these lasing emissions [63]. When these questions are solved, the practical applications of 2D lasers may become eventually realistic.

Controlling the PDOS has also been demonstrated to enhance thermal radiation from high-temperature graphene using Joule heating [28, 64]. Using graphene for broadband thermal emission has many advantages since graphene can reach very high electronic temperatures (>2000K) without significantly heating the underlying substrate [28], which allows thermal emission at visible wavelengths in nanoscale devices [64-66]. Additionally, due to the very low thermal capacitance and high thermal conductivity of graphene, these thermal emitters can be modulated at speeds up to 10 GHz [66]. Shiue et al. recently demonstrated control of thermal emission from a graphene ribbon coupled to a PCC, shown in Figure 2B [28]. By modifying the local PDOS, the measured thermal radiation was highly concentrated at the resonant wavelengths of the PCC. These high-speed, narrow band, and phase-randomized devices could be useful in integrated applications such as discrete-variable quantum key distribution[67] (provided the device is operated in the limit of <1 photon per pulse).

The devices mentioned thus far have been pumped via normal-incidence laser excitation and out-coupled to free space. In order to truly benefit from advances in the field of photonic integrated circuits, it is necessary to couple the light emitted from the 2D material to a neighboring waveguide rather than the far-field. A multifunctional photodetector and LED composed of a $MoTe_2$ split-gate p-n transistor was demonstrated on a silicon photonic crystal waveguide [29]. Depending on the biasing conditions, this p-n diode could be used to either detect or transmit light in the waveguide, enabling the device to act as an integrated transceiver in the near-IR. Recently, a waveguide-coupled single-photon emitter was demonstrated using $WSe_2$, but the coupling efficiency to the waveguide mode was found to be quite weak (between 0% ~ 7.3%, depending on the dipole polarization) [25]. In order to enhance the collection efficiency to the waveguide rather than free space, it is necessary to integrate these devices into waveguide-coupled cavities. However, the coupling between the waveguide and the cavity must be optimized since the coupling to the waveguide decreases the photon lifetime of the cavity (i.e., loaded $Q$ < intrinsic $Q$). Peyskens et al. derived this optimal coupling condition for an integrated cavity-emitter system[25] and found that the extraction efficiency of a single photon emitter could approach unity for Purcell factors on the order of ~$10^3$ when considering low quantum yield emitters ($\eta = 0.01$).

2.3 Plasmonic Structures

Nanoplasmonic resonators allow extreme mode confinement by coupling to coherent oscillations of electrons at a dielectric-metal interface—thus, overcoming the diffraction limit. This extreme reduction in $V_{eff}$ compensates for the lower $Q$ of plasmonic resonators and allows for Purcell enhancement over a broader range of wavelengths than typical dielectric resonators[68]. Enhancement of both absorption and PL emission was achieved, shown in Figure 2G, using a plasmonic nanopatch antenna design[24]. Silver nanocubes scattered on a $MoS_2$ monolayer enabled a plasmonic resonator with extremely small mode volume ($V_{eff} \approx 0.001(\lambda/n)^3$) and a 2000× enhancement of the measured PL was observed.

In addition to PL enhancement, plasmonics can also provide a means for enabling coupling between dark excitons (i.e. excitons with dipole moments normal to the plane of the 2D material) and the far-field, thus providing a method for exploring new physics in these materials directly with an optical probe [69, 70].

Plasmonic metamaterials, which are composed of both metallic and dielectric materials layers with a subwavelength periodicity, can be engineered to exhibit very different optical dispersion relations and therefore highly unique PDOS. To this end, Galfsky et al. showed in 2016 that hyperbolic metamaterials (HMMs) could be used in combination with a PCC to achieve both broadband enhancement of the radiative recombination rates in $MoS_2$ and $WS_2$ in addition to highly directional PL emission (Figure 2H) [55]. In this demonstration, the HMMs increases the PDOS over a very broad range of wavelengths through a phenomenon known as hyperbolic dispersion (i.e., allowed wavevectors are not limited to an elliptical iso-surface dispersion surface, but rather an ideally infinite hyperbolic surface [71]). The use of HMMs allows for Purcell enhancement to occur over a large area and broad range of wavelengths [72-74] compared to resonant photonic structures which seek to maximize $Q/V_{eff}$. However, as noted by Galfsky et al., the majority of photonic modes in the HMMs do not couple to free space, which necessitates the patterning of a periodic photonic crystal in the HMMs. This hybrid "photonic hypercrystal" (PHC) results in Purcell enhancement over a broad range of frequencies, distinct from enhancement at the resonance frequency of a photonic cavity.

Integration of these plasmonic resonators and metamaterials with photonic circuits is arguably more crucial than in the case of dielectric resonators since the propagation loss of plasmonic waveguides is many orders of magnitude greater than that of photonic waveguides. To address this issue, hybrid photonic-plasmonic devices have been successfully demonstrated, which combine the extreme light confinement of plasmonics with the low propagation loss of photonics [75-79]. This hybrid approach has also been used in the context of plasmonic enhancement of 2D materials for both optical modulation [9] and photodetection [33, 80]. Integration of plasmonic structures with photonic circuits and 2D materials could also enable new types of chiral photonic devices which make use of the emitted photon angular momentum in 2D excitonic transitions [81, 82]. The chiral photonic devices will be reviewed in Chapter 4.

## 3. Polaritonics through photonic integration

After the weak coupling regime discussed in the last chapter, we now focus on the strong coupling regime in this chapter. In the strong coupling regime, the interaction between the excitons and cavity photons takes place in the format of reversible energy transfer [40-43]. This means the interaction rate is faster than the average decay rates of the excitons and cavity photons so that energy transfer is back and forth between them. As a result, new eigenstates with prominent anticrossed states called exciton-polaritons (EPs) are formed in the specific energy dispersion diagrams [40-43]. In the simplest example with coupling between one exciton and one cavity mode (Figure 3A), two polariton states, namely upper polariton (UP) and lower polariton (LP), are anticrossed with each other by the Rabi splitting as [40-43].

$$E_{UP,LP} = \frac{E_{cav} + E_{ex}}{2} + i\frac{\gamma_{cav} + \gamma_{ex}}{2} \\ \pm \sqrt{\Omega_0^2 + 1/4[(E_{cav} - E_{ex}) + i(\gamma_{cav} - \gamma_{ex})]^2} \quad (4)$$

where $E_{cav}$ and $E_{ex}$ refer to cavity photon and exciton energies respectively, $\gamma_{cav}$ and $\gamma_{ex}$ as half-linewidths of cavity photon and exciton respectively, $E_{UP,LP}$ as new polariton states, $\Omega_0$ as coupling strength, and the Rabi frequency $\Omega_{Rabi} = 2\sqrt{\Omega_0^2 + [i(\gamma_{cav} - \gamma_{ex})]^2}$ represents the interaction oscillation frequency between the exciton and cavity photon. Based on this equation, the splitting between UP and LP could be observed only when the square root is real as $|\Omega_0| > |\gamma_{ex} - \gamma_{cav}|$, but the strong coupling regime strictly requires that $\Omega_0^2 > \frac{\gamma_{ex}^2 + \gamma_{cav}^2}{2}$ as the interaction rate has to be stronger than the average rates of excitons and photons [40-43, 83, 84].

By taking advantage of the 2D materials, the investigations of strong light-matter coupling in various cavity systems led to extraordinary discoveries and promising applications [34, 38, 39]. As mentioned above, the micro- or nano- cavity systems could not only scale down to the monolayer limit but could also be operated at more practical conditions with greater flexibility due to the large binding energy and huge oscillator strength of 2D excitons [85]. Moreover, these cavity systems could inherit the valley degree of freedom (DoF), which is another fascinating property in TMDs monolayers [86-88] for valleytronics applications [89]. Briefly introduce the valley DoF that will be discussed in detail in chapter 4. The valley DoF in TMDs monolayers stems from the break of reversal symmetry, which results in the energetical degeneration at the extrema of the conduction band in the hexagonal Brillouin zone, which are the so-called K and K' valley [86-88]. Last but not least, the burgeoning progresses of hetero-/homo- multilayers of 2D materials [90-95] prove and prospect the unprecedented possibilities of engineering the 2D material-cavity systems for the profound fundamental physics of cavity quantum electrodynamics (CQED) and more functional applications. In the following section, we discuss the research advancement of these cavities in the strong coupling regime, and we then discuss some novel findings on the heterobilayer-cavities and outlook on the possible future directions.

In the strong coupling regime, EPs could inherit the physical properties of 2D excitons and even manipulate 2D excitons [34, 38, 39]. For the large binding energy and huge oscillator strength of 2D excitons, 2D EPs could be formed with only one single monolayer at various temperatures, including room temperature (RT) [34, 38, 39]. In this context, the strong coupling regime of 2D materials-cavity systems indeed provides a versatile platform to extensively probe the fascinating physical phenomena and engineer the corresponding applications in quantum optics and valleytronics [34, 38, 39].

The strong coupling regime has been widely reached in various 2D materials-Fabry-Perot (FP) cavities with noticeable advances [39]. The first experimental realization, shown in Figure 3B, was carried out at RT in an FP cavity embedded with $MoS_2$ monolayer [16]. An open FP cavity (Figure 3C, where the top mirror could be moved without contacting the cavity layer) was also implemented to demonstrate strong coupling at low temperatures for monolayer $MoSe_2$ [96]. These early experiments determined the strong coupling by meeting the criterion of $\Omega_0^2 > \frac{\gamma_{ex}^2 + \gamma_{cav}^2}{2}$ with parameters extracted from the reflectivity and PL spectroscopy. Since then, various TMDs-FP cavities with metal mirrors or distributed Bragg reflectors (DBRs) have reached the strong coupling regime [39, 97-99]. In the FP cavity configuration, the valley DoF was proved to be well preserved for the strongly coupled EPs and could be even robust at RT (Figure 3D [100]) by various groups [97, 100-103], while this DoF may vanish in excitons at RT for increased intervalley scattering. By optimizing the FP cavity structure, the underlying CQED could be eventually resolved [104] and the nonlinear optics could be established (Figure 3E [105]) to significantly change the valley dynamics of EP excited states and ground states[105, 106]. With controlled valley dynamics by nonlinear optics, a prominent optical valley hall effect was also observed (Figure 3F) [106] by taking advantage of the FP cavity configuration. Moreover, this FP cavity configuration also offers a flexible framework to further explore interesting physics and exploit promising applications. As an example, a gating device could be embedded into the FP cavities to unravel the correlation between the strong coupling and carrier doping[107, 108] and even reveal the strong interactions of polaron-polaritons with excessive carriers (Figure 3G) [108]. In addition, FP cavities could be integrated with heterostructured 2D materials for light-emitting diodes (LED) as Figure 3H [109], opening up the possibilities of electrical pumping for these EPs. Indeed, FP cavities are the most straightforward and convenient option to research the strong coupling of 2D TMDs, and therefore prospect the most possibilities to realize striking phenomena and applications ahead, such as valley polariton Bose-Einstein condensation (BEC), electrically pumped EP laser and valleytronics devices [34, 38, 39].

Plasmonic cavity systems, consisting of metallic nanostructures as the example in Figure 3I [110], have also been employed to look into the strong coupling of 2D materials [110-114]. In these cavities, the mode volume $v$ is generally small, but the $Q$ factor is relatively large. Depending on the specific plasmonic cavities, the criterion $\Omega_0^2 > \frac{\gamma_{ex}^2 + \gamma_{cav}^2}{2}$ cannot always be met in 2D materials for the relatively large $\gamma_{cav}$ determined from the $Q$ factor[115]. To examine this criterion, the parameters of coupling strength ($\Omega_0$) and linewidths ($\gamma_{cav}, \gamma_{ex}$) could be extracted from dark-field scattering spectroscopy. Note here, the valley DoF could be well protected and act as a new information carrier to route different paths depending the symmetry of plasmonic structures [12, 82, 116, 117].

Other photonic structures could also efficiently enter the strong coupling regime for 2D materials as optical cavities do. For instance, a one-dimensional (1D) photonic crystal (PhC) with a grating-like structure has been shown to efficiently couple the 2D excitons with high $Q$ mode into the strong coupling regime at various temperatures as in Figure 3J [118]. DBR, a periodic photonic structure could also support Bloch surface wave (BSW) modes with a high $Q$ factor. EPs have been demonstrated based on the strong coupling between BSW and excitons, as in Figure 3K [119]. These BSW EPs, for the first time, demonstrated strong nonlinear interactions for 2D EPs, which are crucial properties for their possible quantum behaviors. These photonic structures with highly confined modes indicate many new possibilities that are limited by FP cavities, such as delocalized field control, higher emission yields, and more flexible integrations of optoelectronic devices for 2D materials [119].

Besides the research progress of various cavity systems, the emergent heterostructures of 2D materials also suggest an unprecedented potential for a novel research direction in 2D materials photonics [120, 121]. As well-aligned heterostructures could show much higher QY than a sole monolayer, the most convincing lasing behaviors with coherence in 2D materials have been demonstrated as in Figure 3L [120]. By engineering the coupling between 2D heterostructures (such as stacking and twisting 2D materials) and photonic structures, the coupling could also get into the strong coupling regime. When this is realized, the moiré potentials, the stronger nonlinear interactions, and higher QY come into play, implying a revolutionized experimental platform to study novel CQED phenomena that may not have been seen in other material systems.

## 4. Multifunctionality with Metamaterials

In the previous chapters, we have discussed the integration of metamaterials and 2D materials in the weak coupling and strong coupling regime. Metamaterials with subwavelength building blocks, including metallic, semiconductor, or dielectric meta-atoms, have versatile functionalities. The meta-atoms are usually made of plasmonic or dielectric nanoantennas, which are deliberately arranged to allow for direct control of the light phase [122, 123], amplitude [5, 124], and polarizations [125-127]. Metamaterials can not only manipulate the light at the far-field region, such as polarization conversion [126], perfect absorption [128], light modulation [129], and bend light direction [130, 131], but can also manipulate the EM field at the near-field region [132], electrical and magnetic fields distribution controlment, phase control, etc. Layered 2D materials, including allotropes of fundamental elements with hexagonal lattice structures, such as graphene and hexagonal boron nitride (h-BN), and compounds such as TMDs have many exciting and attractive optical properties due to their reduced dimensions, and automatic flatten layered structures [1, 133, 134]. These optical properties, especially in the TMD monolayer, including the valley-selective circular dichroism [86, 87, 135], the valley Hall effect [11], dark-excitons [136, 137], and many others have attracted extended interests of optics and materials societies.

By integrating 2D materials with metamaterials, the light-matter interaction can be enhanced and amplified at the near-field region enabling many enhanced optical functionalities [3]. The polarization of incident light can be converted to other polarized modes in the near-field region. For example, the converted circular polarized light enables the valley-selective circular dichroism in the TMDs monolayer because of the spin-valley locking effects[138, 139]. Due to the spin Hall effect, metamaterials can separate light with opposite chirality towards different directions, which results in the separation of K and K' valley PL signals at the momentum-space (K-space). The manipulation of local electrical field distributions, including their intensity and direction, allows the metamaterials to brighten the dark-excitons [137] in the TMDs monolayers by coupling with the out-of-plane dipole modes of the dark-excitons. The metamaterials can also manipulate the light phase to change the direction of light propagation due to Huygens' principle or the additional geometric phase (Pancharatnam-Berry phase, P-B phase) [140] to achieve a functional super-lens [141]. In this chapter, we will review the multifunctionalities which arise from the integration of 2D materials and metamaterials.

## 4.1 Chiral effects

A material whose molecules or ions are asymmetrical is said to have chirality, a geometric property indicating the lack of mirror symmetry. This term stems from the Greek χείρ (cheir) meaning, the "hand", which is itself a chiral object. The light can interact with the chiral structure of the molecule, becoming rotated and resulting in the macroscopic property known as optical activity. Optical activity was discovered by Arago, who observed the light passed through a quartz crystal placed between two perpendicular polarizers [142]. Following this, optical activity was observed in turpentine by Biot and in solutions of camphor and sugar [142]. However, the optical activity in natural materials is weak and requires a long light path in order to accumulate sufficient chiral effects to facilitate observational accuracy [143]. Metamaterials composed of artificial chiral building blocks of subwavelength size show chiral effects several orders stronger than that in natural materials [144-147]. There are two different types of the chiral metamaterials, plasmonic and Mie resonance [148]. The plasmonic type metamaterials are made of metallic building blocks such as gold, silver, etc., whereas the Mie resonance type metamaterials are made of high index dielectric materials such as silicon nitride ($Si_3N_4$), silicon dioxide ($SiO_2$), vanadium dioxide ($VO_2$), etc., or semiconductors such as silicon (Si) and gallium arsenide (GaAs) [148-152]. Similar to the wave plate, take the quarter-wave plate, for example, introducing the $\pi/2$ phase shift to one of the two perpendicular polarization components of the incident light results in the conversion of a linearly polarized light into circularly polarized light or vice versa, revealing that metamaterials can manipulate light polarization in an efficient way. One of the earliest works to convert linearly polarized light into circularly polarized light with metamaterials was the V-shape plasmonic antenna [130, 153]. Due to the different dipole resonance responses along with two fundamental directions of the V-shape plasmonic antennas, the phase shift can be introduced between two perpendicular polarization components of the scattering light. Recently, many different metamaterial designs have been demonstrated, including the asymmetric split rings [154], anisotropic plasmonic antennas [155], chiral plasmonic structures such as 3D helix structures [156], planars [157, 158], spiral structures [159-161], nanoslits [162]; and dielectric metamaterials such as crossed-bowtie nanoantennas [163], honeycomb structures [164], Mie resonance based asymmetric transmission structures [165], and photonic crystal by breaking the in-plane inversion symmetry (to note that in ref. [166], the authors claimed the broken of C2 symmetry, however, they corrected it as broken of in-plane inversion symmetry in ref. [167]) [126, 166, 167]. To design the metamaterials, for example, the polarization conversion metamaterials, the Jones matrix plays a critical role by linking the incident wave and the desired output wave [168]. For the planar metamaterials, the Jones matrix relates to the complex amplitude of the incident and the scattered fields.

$$\begin{pmatrix} E_r^x \\ E_r^y \end{pmatrix} = \begin{pmatrix} r_{xx} & r_{xy} \\ r_{yx} & r_{yy} \end{pmatrix} \begin{pmatrix} E_i^x \\ E_i^y \end{pmatrix} = R \begin{pmatrix} E_i^x \\ E_i^y \end{pmatrix}$$
$$\begin{pmatrix} E_t^x \\ E_t^y \end{pmatrix} = \begin{pmatrix} t_{xx} & t_{xy} \\ t_{yx} & t_{yy} \end{pmatrix} \begin{pmatrix} E_i^x \\ E_i^y \end{pmatrix} = T \begin{pmatrix} E_i^x \\ E_i^y \end{pmatrix}$$
(5)

where R and T are the reflection and transmission matrices respectively for linear polarization. $E_i^\sigma$, $E_r^\sigma$ and $E_t^\sigma$ are the incident, reflected, and transmitted electric fields polarized along $\sigma$ direction where $\sigma \in \{x, y\}$, respectively. The metamaterials can be designed using the numerical simulation methods, including the finite element methods such as the COMSOL Multiphysics, the finite-difference time-domain method as used in Lumerical and the Meep package [169], the rigorous coupled-wave analysis method [170], and the machine learning or deep learning inverse design method as demonstrated recently [171-173]. To represent the strength of the circular dichroism, the degree of circular polarization (DCP), the Stokes parameters, and the Poincare sphere are the most popular methods. The DCP can be express as $p = \dfrac{I_+ - I_-}{I_+ + I_-}$, where $I_+$ and $I_-$ denote the intensities of light with $\sigma_+$ and $\sigma_-$ polarization, respectively. The stokes parameters, as defined by George Gabriel Stokes in 1852, are a set of values, i.e., $S_0$, $S_1$, $S_2$, and $S_3$, that describe the polarization state of the EM wave [174]. They are shown in Equation 6 in a spherical coordinate system (Figure 4A):

$$\begin{aligned} S_0 &= I \\ S_1 &= I p \cos 2\varphi \cos 2\chi \\ S_2 &= I p \sin 2\varphi \cos 2\chi \\ S_1 &= I p \sin 2\chi \end{aligned}$$
(6)

where the $I$ denotes the intensity of the light, and $p$ is the degree of polarization (DOP). A wave that is partially polarized can be disassembled into the polarized and unpolarized components thus the DOP is represented as the fraction of the polarized power out of the total power of the wave given as a value between 0 and 1. The $2\varphi$ and the $2\chi$ are the azimuthal angles and the polar angle in the Poincaré sphere, the parametrization of the S₁, S₂, and S₃ in spherical coordinates, as shown in Figure 4A. The Stokes parameters can also be vectorized, resulting in the following:

$$\begin{aligned} S_0 &= I = E_x^2 + E_y^2 \\ S_1 &= E_x^2 - E_y^2 \\ S_2 &= 2E_x E_y \cos(\delta_y - \delta_x) \\ S_3 &= 2E_x E_y \sin(\delta_y - \delta_x) \end{aligned} \quad (7)$$

where the $E_x$, $E_y$, $\delta_x$, and $\delta_y$ are the x and y direction polarized electrical field and phase, respectively. It is obvious that (1,1,0,0) and (1,-1,0,0) are representing the linearly polarized light in the horizontal and vertical directions, (1,0,1,0) and (1,0,-1,0) are representing the linearly polarized light in ±45° direction, and (1,0,0,1) and (1,0,0,-1) are the representing the right and left-hand circularly polarized light, respectively.

4.1.1 Valley-selective circular photoluminescence

The valley-selective circular dichroism is one of the most exciting optical properties in TMDs monolayers, as we briefly discussed in chapter 3 [86-88, 135]. These two valleys, at the K and K' points respectively, have opposite Berry curvatures leading to different valley pseudospins. Thus, the K valley responds only to the right-hand circular polarized light that carries the $\sigma_+$ spin angular momentum, and K' responds to the opposite. This is known as the spin-valley-locked light-matter interaction for both absorption and emission of photons. The valley DoF provides an important platform for exploiting new light-matter interaction phenomena such as spintronic devices [11], valley-selective light-emitters [175, 176], and information processing and harvesting [38, 87, 135, 177-180]. To manipulate the valley DoF, metamaterials have been widely employed to control the spin angular momentum of light.

Metal-dielectric-metal plasmonic chiral metamaterials have been shown to control the MoS₂ monolayer far-field exciton emission of a specific valley as shown in Figure 4B [181]. The chemical vapor deposition (CVD)-grown MoS₂ monolayer was put between an LCP chiral metamaterial and a gold reflection layer, which contained a plasmonic chiral microcavity. The LCP chiral structure enhanced the absolute value of the degree of valley polarization (DVP), defined as $p=\frac{I(\sigma_+)-I(\sigma_-)}{I(\sigma_+)+I(\sigma_-)}$, from 25% ± 2% to 43% ± 2% at 87k under the excited LCP light ($\sigma_-$) and suppressed the absolute value of DVP to 20% ± 2% under the excited RCP light ($\sigma_+$). The ability to identify valley-polarized PL is indicative of a bright future in the development of valley-dependent optoelectronic devices. In parallel, a WSe$_2$ monolayer used between the chiral plasmonic structures and silicon dioxide substrate has demonstrated superb valley-polarized PL control, shown in Figure 4C [182]. The phonon-assisted intervalley scattering sharply accelerated when the temperature increased, resulting in significantly reduced usability of the far-field PL signal, especially at room temperature [87, 183, 184]. The optical cavities have been widely used for removing the cryogenic-temperature restriction thus making the room-temperature valley dynamics be controllable [97, 100-102, 185]. A tunable moiré chiral metamaterial (MCM), as shown in Figure 4D, is made of two layers with a gold nanohole hexagon array, which allows for twist tunability for the interlayer in-plane angle. This takes advantage of the so-called chiral Purcell effect to modulate the valley dynamic of a monolayer WSe$_2$ [186]. Similar to the Purcell effect, the enhancement of the spontaneous emission rates of atoms that incorporated into a resonant cavity discovered by Edward Mills Purcell in the 1940s, the chiral Purcell effect shows that the chiral metamaterials or chiral environments can modify the spontaneous emission rates of valley excitons at the K and K' valleys ($\Gamma_+ = 1/\tau_+$ and $\Gamma_- = 1/\tau_-$, respectively) recombine and then emit the $\sigma_+$ and $\sigma_-$ circularly polarized photons. The MCMs can selectively enhance the spontaneous emission rate of valley excitons at K valley yielding as large as 60% of degree of circular polarization ($DCP \approx \frac{\Gamma_+ - \Gamma_-}{\Gamma_+ + \Gamma_-}$) and of valley excitons at K' valley with -60% DCP.

4.1.2 Valley-hall effects

The Hall effect, discovered by Edwin Hall in 1879, illustrates the phenomenon of a voltage bias (Hall voltage), which is observable across an electrical conductor, transverse to an electric current in the conductor when a magnetic field is applied perpendicular to the conductor[187]. The effect originates from the Lorentz force with opposite signs applied to the electrons or holes, driving them in divergent directions. After the Hall effect's discovery, a family of Hall effects including the anomalous Hall effect, the quantum Hall effect, the spin Hall effect (SHE), the quantum SHE, the quantum anomalous Hall effect, and the valley Hall effect have been observed [188-196]. Most of these effects originate from the Lorenz force or quasi-Lorenz force due to the perpendicular magnetic field or quasi-magnetic field [197]. The Hall voltage can be observed even without the external magnetic field, i.e., the anomalous Hall effect, because the intrinsic magnetic field of the material itself can provide the Lorentz force [188]. Further, the Hall voltage can be observed without either external or intrinsic magnetic fields known as the spin Hall effect [190-192]. Spin, another degree of freedom of electrons, with two possible signs, can be treated as a magnetic moment along the spin direction. As its optical counterpart, the optical spin Hall effect manifests itself as spin-dependent shifts or spin accumulation of photons [193, 194]. The geometry phase (Berry phase) acts as the quasi-magnetic field in the photonic spin Hall effect [140]. The Berry phase in optics can be interpreted as the coupling between the spin angular momentum (SAM) of a photon and coordinate rotation. The early optical SHE was investigated in total reflection and gradient-index materials, namely the Imbert-Fedorov effect [198, 199] and the optical Magnus effect [200, 201], respectively. The optical SHE was linked to the geometric Berry phase and conservation of angular momentum by Onoda et al. in 2004 [202]. Soon after, the complete theoretical model of the optical SHE regarding the refraction/reflection of paraxial light beams with inhomogeneous media interfaces was built [203, 204]. This model was verified via weak measurements about two years later [205]. Figure 5A shows the schematic of the optical SHE in the diffraction of an oblique incident light beam with an incident angle $\theta$ and right circular polarization from air to glass[205]. The beam of light can be bent at the interface of air and glass because of the refractive index difference of those two mediums. The total angular momentum in the z-direction ($J_z$, the sum of spin-angular momentum and trajectory angular momentum) must be a constant due to the symmetry (angular momentum in the z direction conservation). Take the RCP of light for example, the spin-angular momentum of the refracted light should be larger than that of incident light due to its smaller propagation angle. As a result, the external orbital-angular momentum with the same absolute intensity but with opposite sign (+ or -) should originate to cause the change of spin-angular momentum along z direction, which yields movement of the photon towards the y direction. Similar results occur with LCP light but with photon movement towards the opposite direction. Summarily, the optical SHE stems from the interaction between the spin-angular momentum and the external orbital-angular momentum.

The optical SHE illustrates that linearly polarized light can be separated into two beams with opposing spin signs, i.e., RCP and LCP lights propagate towards divergent directions. When integrated with the TMDs, the K and K' valley excitons can be excited spatially separated. Taking advances from the optical SHE, a silver nanowire (AgNW) sited on a $WS_2$ thin flake can spatially separate the RCP and LCP light and excite the K and K' valley PL of the $WS_2$ flake as shown in Figure 5B [12]. Linearly polarized light can be coupled into surface plasmon polaritons (SPPs) modes by the AgNW and propagate towards the two ends of the nanowire [206]. RCP and LCP light, however, would experience a slight shift to the right and left and be spatially separated to the opposite side of the AgNW, as illustrated by Figure 5B (top right). When the coupling point of the incident light is not at the center of the nanowire, for example when y>0, this leads to higher coupling efficiency on the y>0 side that comparing to the opposite side of the AgNW. As a result, more RCP light goes to $-k_x$ direction and LCP light goes in a divergent way. The spatially separated light with varying circular polarizations excited different valley excitons and yielded a different valley-selective PL at each side, as shown in Figure 5B (lower right).

Although the optical SHE is a weak phenomenon in the nature [205], it can be largely amplified by the metasurfaces [207-210]. Akin to the AgNW configuration, an asymmetric groove array manifested itself with the feasibility to sort and spatially separate the valley-polarized excitons of a $MoS_2$ monolayer with different chirality [117]. The symmetric groove array, as shown in Figure 5C (left), supports both RCP and LCP light for each propagating direction ($k_y$ and $-k_y$). By breaking the mirror symmetry, i.e., using the asymmetric design, the RCP and LCP are prevented from propagating toward the $k_y$ and $-k_y$ directions, respectively. As a result, the circularly polarized light with different chirality can be separated spatially along the y-direction and excite K or K' valley excitons of the $MoS_2$ monolayer that are integrated at the top of the groove array.

In addition to the dielectric metasurface, a gold metasurface, as shown in Figure 5D, consists of an array of rectangular nanoslits with in-plane rotation angle $\theta$, yielding a birefringent wave plate [211]. For example, RCP light incident to the gold metasurface would split into two components due to the spin-orbit coupling. The first component is an LCP beam with a geometry phase shift $e^{i2\theta}$, and the other is the RCP residue beam, which maintains the same geometric phase as the incident light. Those two beams can be separated at the K-space due to their different geometric phases and can interact with the $WS_2$ monolayer integrated on the top of the gold metasurface, leading to a second harmonic generation (SHG) signal with a different circular polarization. The LCP beam with a geometric phase shift $e^{i2\theta}$ would result in an RCP SHG signal with a $e^{i4\theta}$ shift, where the RCP residual beam generates an LCP SHG signal without a geometric phase shift, as shown in Figure 5D (middle). The sign of the circular polarization of the SHG signal originates from the so-called nonlinear selection rule [212, 213]. Similar work was reported shortly after [214].

An all-dielectric photonic crystal (PhC) with in-plane inversion symmetry breaking building blocks, i.e., triangular holes, demonstrates circular polarization conversion ability at the vicinity of the bound state in the continuum (BIC), as shown in Figure 3E [166, 167]. The RCP and LCP light can be separated in the K-space (Figure 5E (lower right)), leading to the spatial separation of the lights based on their different circular polarizations.

4.2 Light phase engineering

The phase is another fundamental property of the light. The phase is an angle that denotes the number of periods of a periodic function (such as the EM wave). It can be measured in degrees or radians and thus 360 degrees or $2\pi$ represents a full period. The Huygens-Fresnel principle, named after Dutch physicist Christiaan Huygens and French physicist Augustin-Jean Fresnel, is one of the most famous analysis methods that uses the phase of the light[215]. This principle states that every point on a wavefront is the source of spherical wavelets that have the same frequency and speed of propagation as the point from which it originates. The sum of the spherical wavelets with the same phase forms the wavefront, which dictates the direction of wave propagation. Figure 6A and B present the two fundamental phenomena in the field of linear-optics analyzed by the Huygens-Fresnel principle, namely refraction and diffraction, respectively. In Figure 6A and B, the blue lines represent the wavefront before refraction or diffraction, while the green lines show the wavefront after refraction or diffraction, and the yellow dots denote the sources of spherical wavelets, all while assuming the medium interface is inhomogeneous. Those new sources of spherical wavelets at the interface plane have a different phase because of the oblique incident light. Take the refraction; for example, one can easily draw the refracted direction by finding the plane with the same phase of those sources. The artificial plasmonic or dielectric metamaterials or metasurfaces can adjust the light phase in the subwavelength scale which results in the manipulation of the light propagating direction [216, 217].

The first optical metasurfaces were made of periodic resonant gold antenna arrays on silicon oxide substrates, as shown in Figure 6C, to bend the incident light beam by manipulating the light phase [130, 218]. The electric field of the incident light can be projected to two orthometric directions along and perpendicular to the axis of mirror symmetry of the V-shape antenna. Varying angles of the V-shape antenna achieves different combinations of those electrical field components allowing for precise phase manipulation. However, the Ohmic loss in the metallic nanostructures would restrict the performance of the metasurfaces [217]. As an alternative option, all-dielectric metasurfaces have emerged rapidly and bloomed in recent years [216]. Dielectric nanostructures or nanoparticles as building blocks for the metasurfaces can exhibit strong Mie-type resonances of both the electric and magnetic fields depending on their size. At resonances, either the electric or magnetic dipole or quadrupole or even higher-order momentums would be supported by the building blocks and yield strong scattering. As a result, the transmitted light of the metasurfaces would experience a phase delay in the full range of $0$-$2\pi$. Some experimental examples have been successful in various operating wavelength ranges, from the ultraviolet to the mid-infrared. Typically, $TiO_2$ [219, 220], GaN [221, 222], Silicon [223-225], and $Si_3N_4$ [226, 227] usually serve as the building blocks for these dielectric metasurfaces. The all-dielectric metasurfaces have achieved high efficiencies such as 86% at a wavelength of 405nm using $TiO_2$ [131], 92% at 532nm using GaN [222], 60% at 266nm (ultraviolet) using $HfO_2$ [228], 70% and 82% at 850nm and 1550nm using amorphous silicon, respectively[229, 230], etc. Take the $TiO_2$ nanofin design as an example (Figure 6D), the length and width of each nanofin (in the $TiO_2$ nanofin array) are 250nm and 80nm, respectively [131]. Figure 6E shows an FDTD simulation of the effective refractive index of the nanofin [216]. Polarization of the light along the long or short axis of the fin can support different Mie-type resonances and leads to different effective refractive index. In Figure 6E the red curve shows the effective index for the light polarized along the long axis while the blue curve is polarized along the short axis. The anisotropic behavior of this structure is similar to that of the waveplate and can achieve phase manipulation. The design uses the anisotropic structures as the building blocks, and the transmitted phase delay depends on the nanofin's twist angle.

One of the most interesting applications of metasurfaces are metalenses [231-233]. Like the conventional optical lens, a beam of parallel transmitted light can be bent and focused on its focal plane. The total optical path of each portion of the beam can maintain a constant because of the lens profile. Based on the same fundamental principle, precisely designed metasurfaces with different phase delay distributions can mimic a lens. To design a metasurface, the first step is to calculate the target phase delay distribution and the equation (8) for a focal lens:

$$\Phi(r) = -\frac{2\pi}{\lambda_d}(\sqrt{r^2 + f^2} - f)$$

$$r = \sqrt{x^2 + y^2}$$

(8)

where $x$ and $y$ are the radial coordinates, and $f$ is the focal length, $\Phi$ denotes the phase delay distribution to make compensation to mimic a lens. The next step is choosing a proper metasurfaces design and calculating the building blocks distribution. There are many designs for a metalens such as the nanofins (Figure 6D) [216], nanopillars [228], the so-called "dielectric gradient metasurfaces optical element" (DGMOE) using the PB phase (Figure 6F) [234], etc.

Integrating 2D materials with phase control metasurfaces has resulted in several applications, including the nonlinear metalens [235], unidirectional PL emission [236, 237], etc. The gold film was patterned by FIB that formed a metalens has manifested itself with feasibility for enhancing and focusing the SHG signal that excited from the integrated $WS_2$ monolayer on top, as shown in Figure 6G [235]. In parallel, all-dielectric photonic crystals can selectively enhance the PL signal and change the signal transmitted direction, as shown in Figure 6H [236]. The transmitted direction can be designed by calculating the dispersion relationship (Bandstructures). Similar work was reported soon after [237]. Recently, several works reported that 2D materials, including graphene, TMDs, h-BN, etc., can be patterned into the periodic structures and worked like the metalenses directly [238, 239], which would be reviewed in Chapter 5.

4.3 Other functionalities

In addition to the aforementioned functionalities of metamaterials or metasurfaces reviewed above, near-field EM field tailoring, nano-lasing, graphene doping, and many other functions will be reviewed in this section.

Using normal incident light, regardless of the polarization or even if non-polarized, it is hard to achieve out-of-plane electric or magnetic dipole excited. However, out-of-plane electric or magnetic dipole momentums are important to facilitate observation of some optical phenomena including dark-exciton brightening [136, 137, 240], out-of-plane Raman vibrating modes pumping [206, 241], etc. The dark excitons ($X_D$) have a long lifetime and have potential applications in the field of quantum information processing [242], or Bose-Einstein condensation (BEC) [243]. However, their out-of-plane transition dipole momentums make their optical access difficult [136, 137, 240, 244]. Silver nano grating can couple the normal incident laser into SPPs modes, and in conjunction with silver surfaces can lead to an out-of-plane electric dipole. This design has been applied to brighten the $WSe_2$ and $MoSe_2$ monolayer dark excitons [137]. However, the absence of the electric field enhancement makes this dark exciton brightening limited to cryogenic temperature conditions (~4 K), which limits the scope of application. An all-dielectric PhC nanocavity has been designed to work on TM-like modes (the even mode, E-field distribution maintain Z-symmetry) and been demonstrated the out-of-plane electric field converter leading to dark exciton brightening and PL signal enhancement a [245, 246].

The plasmonic metasurfaces can serve as an antenna to support the weak coupling between a spectrally broad bright plasmon mode and a spectrally narrow dark mode to form a Fano resonance, leading to reduced scattering and enhanced absorption of incident light [247]. Different designs of gold metasurfaces have demonstrated the feasibility of using induced negative or positive charge to inject into the graphene monolayer beneath the metasurfaces. Those injected charges tune the Fermi level of the graphene monolayer and result in the n- or p-type doping of the graphene. Tuning the Fermi level provides a new degree of freedom to exploit the extraordinary properties of the graphene monolayers, such as Pauli blocking. When the Fermi level of charged graphene is high enough, the graphene loses the ability to absorb relative broadband light in the proper energy range [248]. Plasmonic metasurfaces have been placed on top of graphene monolayers in order to tune its Fermi level.

In this chapter, we have reviewed the multifunctionality of metamaterials and metasurfaces, including chiral generating and enhancing, light phase engineering, near-field polarization tailoring, etc. detailing the integration of 2D materials with functional metamaterials and metasurfaces.

## 5. Metasurfaces patterned on 2D materials

2D materials are composed of a layered crystal structure with strong in-plane bonds, where layers are coupled together by weak vdW forces. The abundance and the rich variety of 2D materials enable use with a broad bandgap range, which spans across the electromagnetic spectrum from terahertz, infrared, and visible to ultraviolet. Therefore, it can meet the demanding requirements of various operating frequencies, differentiating it from the conventional metal and dielectric materials [249]. As a result, directly patterning 2D materials to form metamaterials or metasurfaces is paving a new way to explore more functionalities.

Graphene is a promising 2D carbon material, whose electrons behave like massless Dirac fermions as well as display many exotic physical phenomena; in addition, graphene represents an attractive alternative to metamaterials due to its strong field confinement and versatile tunability [250-252]. Wang et al. reported plasmon excitation within the THz frequency range micro-ribbon arrays by fabricating graphene, as shown in Figure 7A, in 2011 [253]. The graphene plasmon resonances can be adjusted in a wide frequency range by changing electrostatic doping as well as the ribbon width. Further, the graphene plasmon frequency also indicates a power-law behavior characteristic, which has been confirmed in 2D massless Dirac electrons. At the plasmon resonance frequency, the graphene plasmons couple to terahertz radiation yielding over 13% for absorption at room temperature in contrast to the plasmon absorption peaks that were observed only in traditional 2D electron gas at low temperature. In 2013, the patterned graphene nano-resonator array manifested itself with the ability to highly confine the tunable energy in the mid-infrared region by Brar et al. [254]. As shown in Figure 7B, they observed the modes of tunable plasmonic graphene nano-resonator array by infrared microscopy. And then, they probed graphene plasmons with $\lambda_p \leq \lambda_0/100$ and plasmon resonances as high as 310 meV (2500 cm$^{-1}$) by changing resonator width and in situ back-gating. Besides, electromagnetic calculations verified graphene plasmonic modes could strongly increase light−matter interactions, for which the local density of optical states is more than $10^6$ larger than that in free space. In 2018, Guo et al. reported a mid-infrared graphene detector, which can monitor a significant temperature dependence of carrier transport by means of localizing barriers associated with a disorder in the nanoribbons, and the plasmon decay in the vicinity of graphene resonators can be detected by electrical methods [255]. Figure 7C depicts the structure of graphene metasurfaces photodetector using standard nanofabrication methods, which consist of multiple graphene-disk plasmonic resonators (GDPRs) and graphene nanoribbons (GNRs). The resonance peak can be shifted to the THz spectral range by changing the shape and size of GDPRs.

In addition to graphene, some other 2D materials have been patterned into metamaterials, such as h-BN, TMDs, etc. The single crystal h-BN is an anisotropic layered material, where $\varepsilon_x = \varepsilon_y = \varepsilon_\perp \neq \varepsilon_\parallel = \varepsilon_z$ [256]. Further, the isotropic in-plane permittivity can be broken by patterning the h-BN flake. Figure 7D (left) shows the schematic of a strategy that patterns a 20nm thickness h-BN flake into a periodic nanoribbon array, with ribbon width of $w$ and gap $g$ between each ribbon [256]. The effective permittivities were given as:

$$\varepsilon_{eff,x} = 1 / \left[ \frac{1-\zeta}{\varepsilon_{h-BN,\perp}} + \zeta \right]$$
$$\varepsilon_{eff,y} = \varepsilon_{h-BN,\perp}(1-\zeta) + \zeta \qquad (9)$$
$$\varepsilon_{eff,z} = \varepsilon_{h-BN,\parallel}(1-\zeta) + \zeta$$

where $\zeta = g/(w+g)$ is the so-called filling factor.

The permittivities along three axes (x, y, and z) can be tuned using the varying filling factor. For example, $\varepsilon_{eff,x} = 3.7$, $\varepsilon_{eff,y} = 15.2 + 0.6i$, and $\varepsilon_{eff,z} = 2.1$ can be obtained when w=70nm and g=30nm at the mid-infrared region (1425cm$^{-1}$). Using the numerical simulation, the electrical field distribution in the real- and momentum-space can be calculated by means of placing an out-of-plane electric dipole at the top surface of the h-BN flake or h-BN metasurface, which mimics an NSOM probe excited z-polarized electric dipole process at the gap between the probe apex and sample surface[257, 258]. Figure 7D (right) shows the comparison of the real part of the E$_z$ distribution at the K-space between h-BN flake and h-BN metasurface, which implies the strong in-plane anisotropic properties of the patterned h-BN. Besides the work that is taking advantage of anisotropic permittivities of h-BN, h-BN flake has been patterned into nanopillars array and manifested itself as a metalens. The schematic of this application is shown in Figure 7F (left), and the optical image of the h-BN metalenses working at the wavelength of 450nm is shown in Figure 7F (right) [238].

Other than the plasmonic and Mie resonance in metallic and dielectric materials, which are essential building block materials for conventional metasurfaces, excitonic resonance has manifested itself to have the ability to build a metasurface with resonance energy tunability by applying varying gating voltages. Recently, Greop et al. reported a carved monolayer WS$_2$ metalens that can focus transmitted light with 0.08% efficiency, as shown in the schematic in Figure 7E (left) [239]. In addition, the gating voltage of the WS$_2$ device can be applied through the ionic liquid system. By tuning the gating voltage from 0V to 3V, the neutral excitons (B and A) can be reduced while the charged trion (A$^-$) can be excited and at resonance, leading to the wavelength tunability of the metalens. It is important to note that the metalens is made of patterned chemical vapor deposition (CVD) grown WS$_2$ monolayer whose quality is worse than its exfoliated counterpart, which reduced the totally focal efficiency. Further, the poor light-matter interaction of monolayer WS$_2$ with around 0.6nm thickness is another reason which impacted the efficiency. There are two possible ways, as perspectives, to enhance the efficiency of the excitonic resonances TMDs monolayer metasurfaces. The first is to get the help from those relative mature plasmonic or Mie resonances metasurfaces, by integrating those TMDs monolayer structures with conventional metasurfaces. Note that the metallic surface may lead to quenching effects of the in-plane excitonic dipoles. The second method takes advantage of the large second-order nonlinear optical susceptibility, $\chi(2)$, which stems from the breaking of reversal symmetry in TMDs monolayers. Second-harmonic generation (SHG) signals ($2\omega_0$) that are generated by TMDs monolayer can be designed to focus on the focal plane while the original light ($\omega_0$) can be filtered by an edge filter.

## 6. Conclusions

To summarize, we briefly discussed the optical properties and functionalities of 2D materials in the introduction. Then, we reviewed the engineering of optical environments in four aspects: enhancement at the weak coupling regime in Chapter 2, confinement in the strong coupling regime in Chapter 3, multifunctionalities in Chapter 4, and 2D materials itself patterning in Chapter 5. In addition, we made some perspectives, such as exploring new strategies for light confinement in the strong coupling regime by means of stacking or twisting 2D materials in Chapter 3, and two ways to enhance the TMDs monolayer metalens efficiency in Chapter 5.

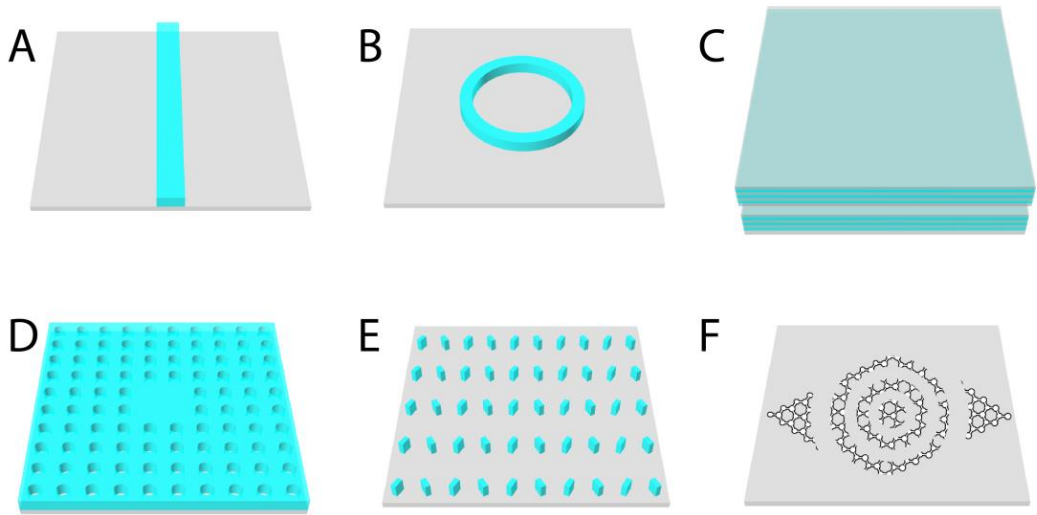

**Figure 1:** Photonic environments for shaping light-matter interactions in 2D materials. (A) Waveguide supports broadband optical responses. (B) Micro-resonator possesses a high quality-factor. (C) Two distributed Bragg reflectors (DBRs) form a cavity in between. (D) A defect in a photonic crystal can also be a cavity. (E) Optical metamaterials design electromagnetic properties at will, supporting multi-functionalities. (F) Electronic metamaterials use 2D materials as building blocks. The figure shows an example of such electronic metamaterials by direct patterning.

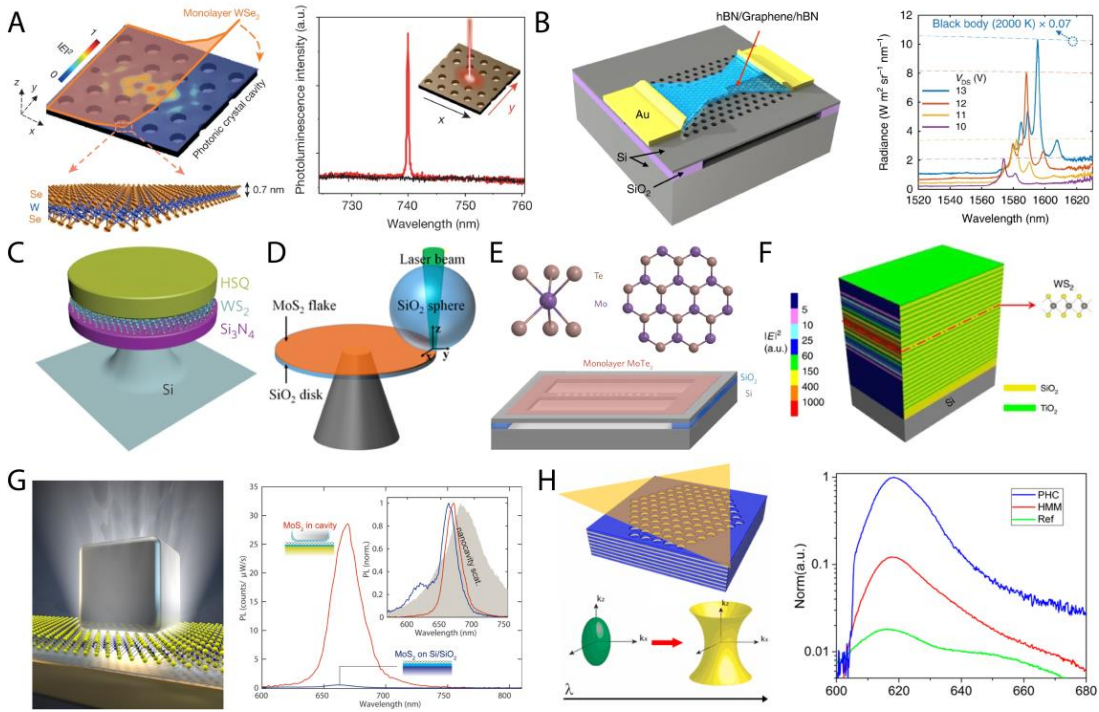

**Figure 2:** Enhancing optical emission by controlling the photonic density of states (PDOS).
(A) Nanocavity laser comprised of monolayer WSe$_2$ on a photonic crystal cavity (PCC). The lasing threshold is reduced by enhancing the Purcell factor [56]. (B) Controlling the thermal emission spectrum of a graphene nanoribbon using a PCC enables narrowband emission at the cavity's resonant wavelengths [28]. (C) The sample schematic for a monolayer WS$_2$ laser based microdisk cavity. The quality factor is around 2600 [58]. (D) The schematic for the preserved valley polarization in strongly coupled exciton polaritons. The quality factor is also between 2600-3300 [60]. (E) The sample structure for a monolayer MoTe$_2$ room temperature CW laser based on a nanobeam cavity [61]. (F) The schematic for a monolayer WS$_2$ room temperature laser based VCSEL [59]. (G) An ultralow mode-volume plasmonic resonator is used to both enhance resonant absorption and photoluminescent efficiency. The resulting enhancement in the photoluminescent yield is approximately 2000x [24]. (H) A hybrid "photonic hypercrystal" uses a hyperbolic metamaterial to increase the PDOS and a photonic crystal to control out-of-plane emission resulting in very broadband and directional enhancement to the photoluminescent signal [55].

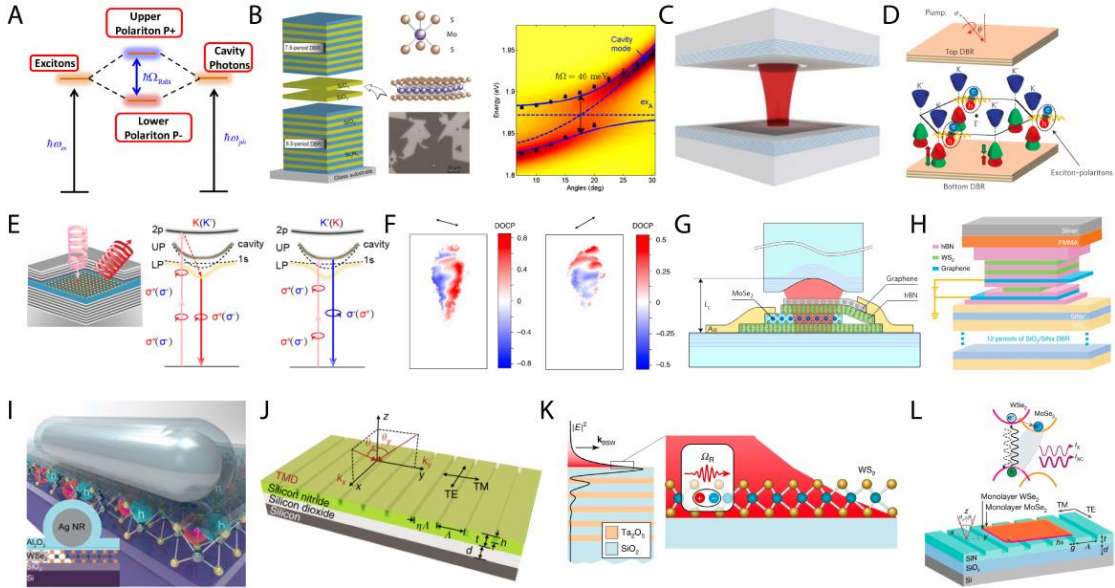

**Figure 3:** The strong coupling regime in 2D materials-cavity systems.
(A) The formation of two polariton states (upper and lower polaritons) in the strong coupling regime of one exciton state and one cavity mode. (B) The first demonstration of strong coupling regime for monolayer $MoS_2$ at room temperature in a FP cavity [16]. (C) The schematic of an open FP cavity to demonstrate the strong coupling regime in monolayer $MoSe_2$. The top DBR can be moved vertically with a nano-translation stage [96]. (D) The schematic for the preserved valley polarization in strongly coupled exciton polaritons [100]. (E) The schematic for the established nonlinear optics to probe the dynamics of excited states of exciton polaritons [105]. (F) The optical valley hall effect could separate the propagations of exciton polaritons into two valley polarized paths as shown in the data via the nonlinear optical excitation [106]. (G) The schematic for the sample structure to study the strong interactions in polaron polaritons [108]. (H) The schematic for the sample device to realize the polariton LED in a monolayer $WS_2$ [109]. (I) A typical example of plasmonic cavity samples for the strong coupling in a monolayer $WSe_2$ [110]. (J) The schematic for a one-dimensional crystal to realize the strong coupling with monolayer TMD [118]. (K) The schematic of Bloch surface wave (BSW) to strongly couple with monolayer $WS_2$ excitons for nonlinear interactions [119]. (L) The schematic of a photonic crystal to realize lasing based on interlayer excitons in a $MoSe_2$-$WSe_2$ heterostructure [120].

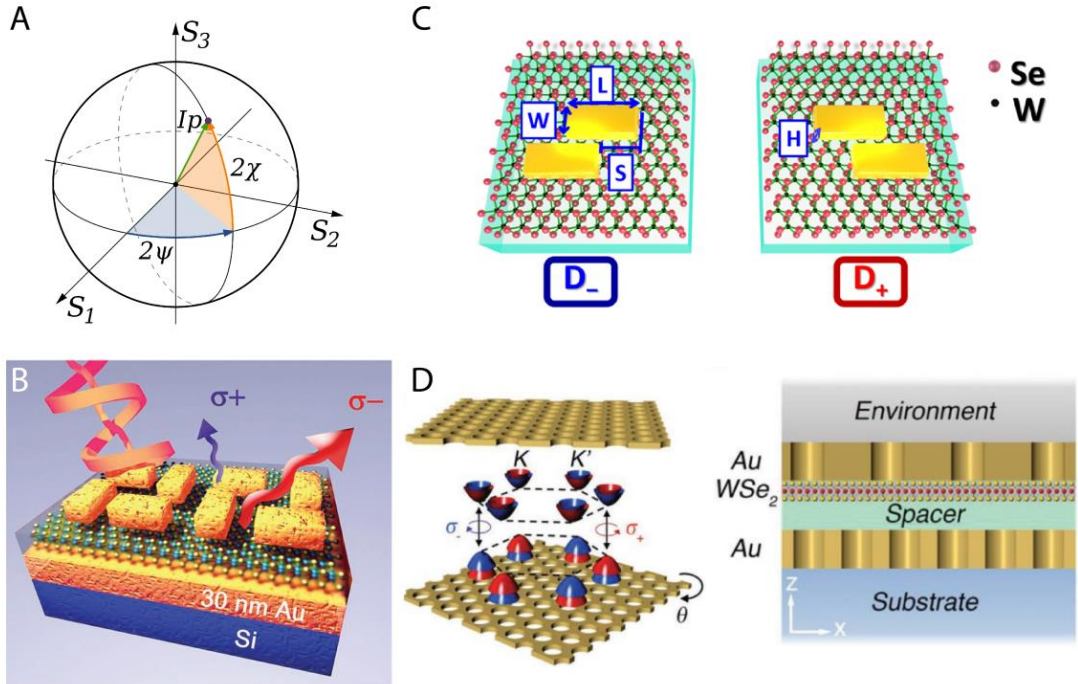

**Figure 4:** Controlment of the valley-selective circular dichroism using chiral metamaterials.
(A) The Poincaré sphere, the parametrization of the S1, S2, and S3 in spherical coordinates. (B) The metal-dielectric-metal plasmonic chiral metamaterials: the CVD-grown $MoS_2$ monolayer was sandwiched between gold substrate and an LCP chiral metamaterial. The k' valley of the $MoS_2$ was enhanced and led to the enhancement of the degree of valley polarization (DVP) from 25%± 2% to 43% ± 2% under the LCP incident light and suppressed the (DVP) to 20% ± 2% under the RCP incident light at 87k [181]. (C) A $WSe_2$ monolayer was sandwiched between dielectric substrate and gold chiral metamaterials, the superb valley-polarized PL has been demonstrated [182]. (D) A tunable moiré chiral metamaterials (MCMs) is made of two layers with a gold nanohole hexagon array and a dielectric spacer layer between those gold structures. A $WSe_2$ monolayer was inserted in the MCMs structure has demonstrated the chiral Purcell effect at room temperature [186].

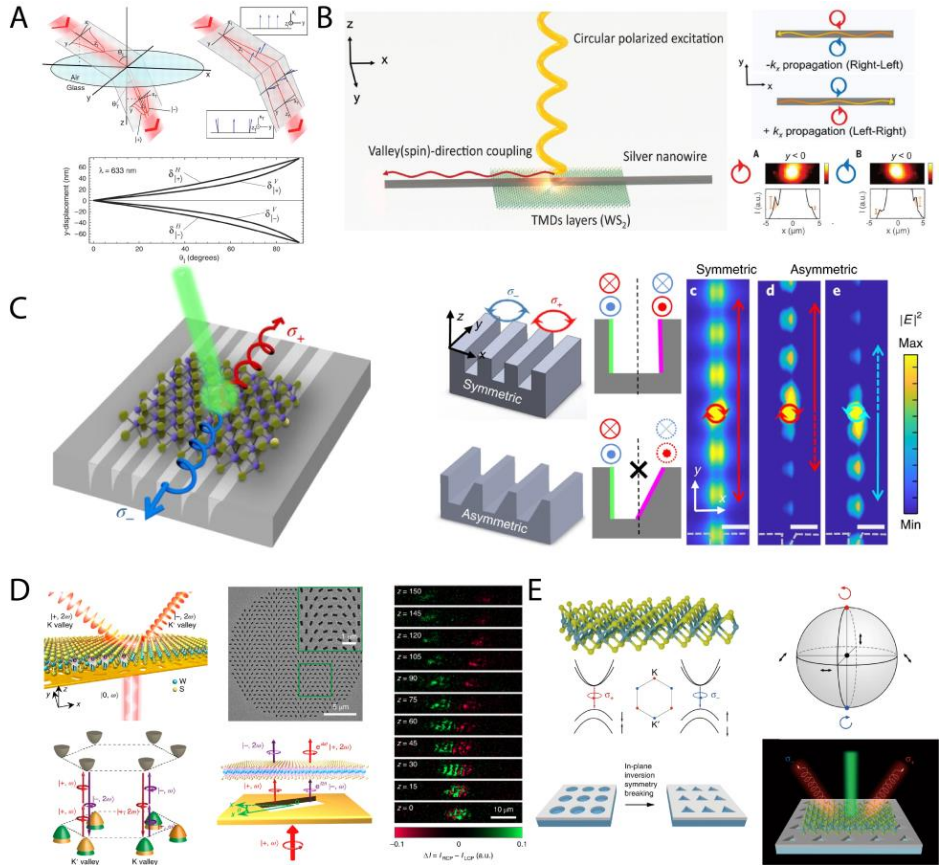

**Figure 5:** Metamaterials separate linearly polarized light into RCP and LCP light due to the optical spin Hall effect.
(A) the schematic of the optical spin Hall effect (SHE) in the diffraction of an oblique incident light beam with an incident angle θ and right circular polarization from air to glass. Photons with $\sigma_+$ or $\sigma_-$ spin angular momentum acquired the momentum along y or -y direction at the interface of the two mediums, respectively [205]. (B) Linearly polarized incident light can be separated into RCP and LCP light by a silver nanowire (AgNW) (left). SPPs modes induced by the AgNW propagate towards two directions: the separated RCP light at right side and LCP light at left side, respectively (top right). The WS$_2$ monolayer beneath the AgNW can be excited the K or K' valley-selective PL corresponding to the separated RCP or LCP light (lower right) [12]. (C) Linearly polarized incident light can be separated into RCP and LCP light due to the optical SHE. An asymmetric grating supports propagating SPPs modes at its cliffy sidewall rather than its oblique sidewall. The RCP and LCP light can be spatially separated towards divergent directions [117]. (D) A gold metamaterial can separate linearly polarized transmitted light into RCP and LCP light at the momentum-space. The K and K' valley selective PL of a monolayer WS$_2$ that is on the metamaterial would be excited and separated [211]. (E) An in-plane inversion symmetry breaking photonic crystal has manifested itself have the RCP and LCP light separation ability [166].

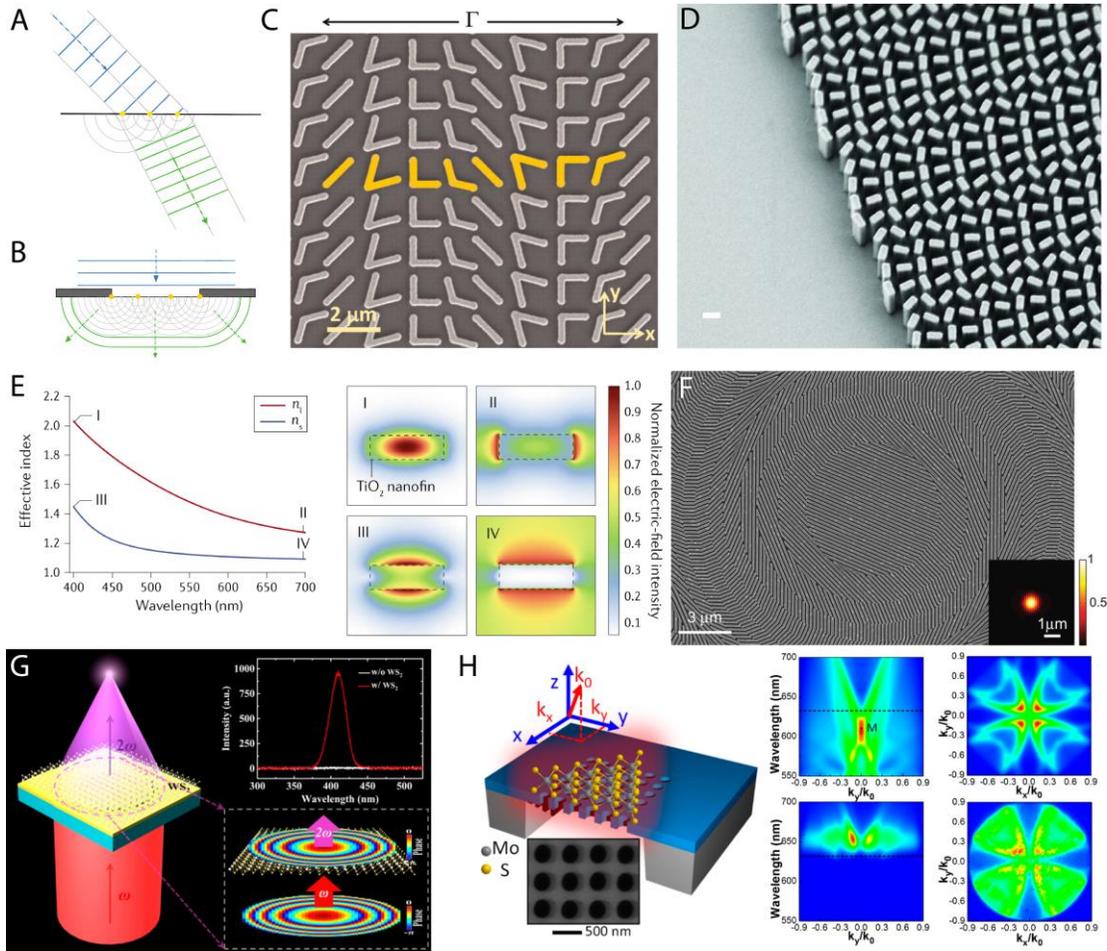

**Figure 6:** Phase engineering of metamaterials integrates with 2D materials.
(A) and (B) Two fundamental phenomena in the field of linear-optics analyzed by the Huygens-Fresnel principle, namely refraction and diffraction, respectively. (C) The V-shape gold antenna can control the phase of transmitted light. Varying the angle of the V-shape antenna allows for precise phase manipulation [130]. (D) The $TiO_2$ nanofin array with varying angle can delay different phase of the transmitted light correspondingly. The SEM image of the fabricated metalens with scale bar of 300nm [131]. (E) A FDTD simulation of the effective refractive index of the nanofin. Polarization of the light along long or short axis of the fin can be supported by the fin in different modes and leads to different effective refractive index [216]. (F) The so-called "dielectric gradient metasurfaces optical element" (DGMOE) using the PB phase designed. The pattern has manifested itself as the metalens that tightly focused the Incident light into a small volume (~1μm) [234]. (G) A metalens can enhance and focus the second harmonic generation (SHG) signal of a $WS_2$ monolayer that is on its top [235]. (H) A suspended photonic crystal (PhC) wavelength selectively enhanced the PL signal of a $MoS_2$ monolayer on its top and changed the signal transmitted direction [236].

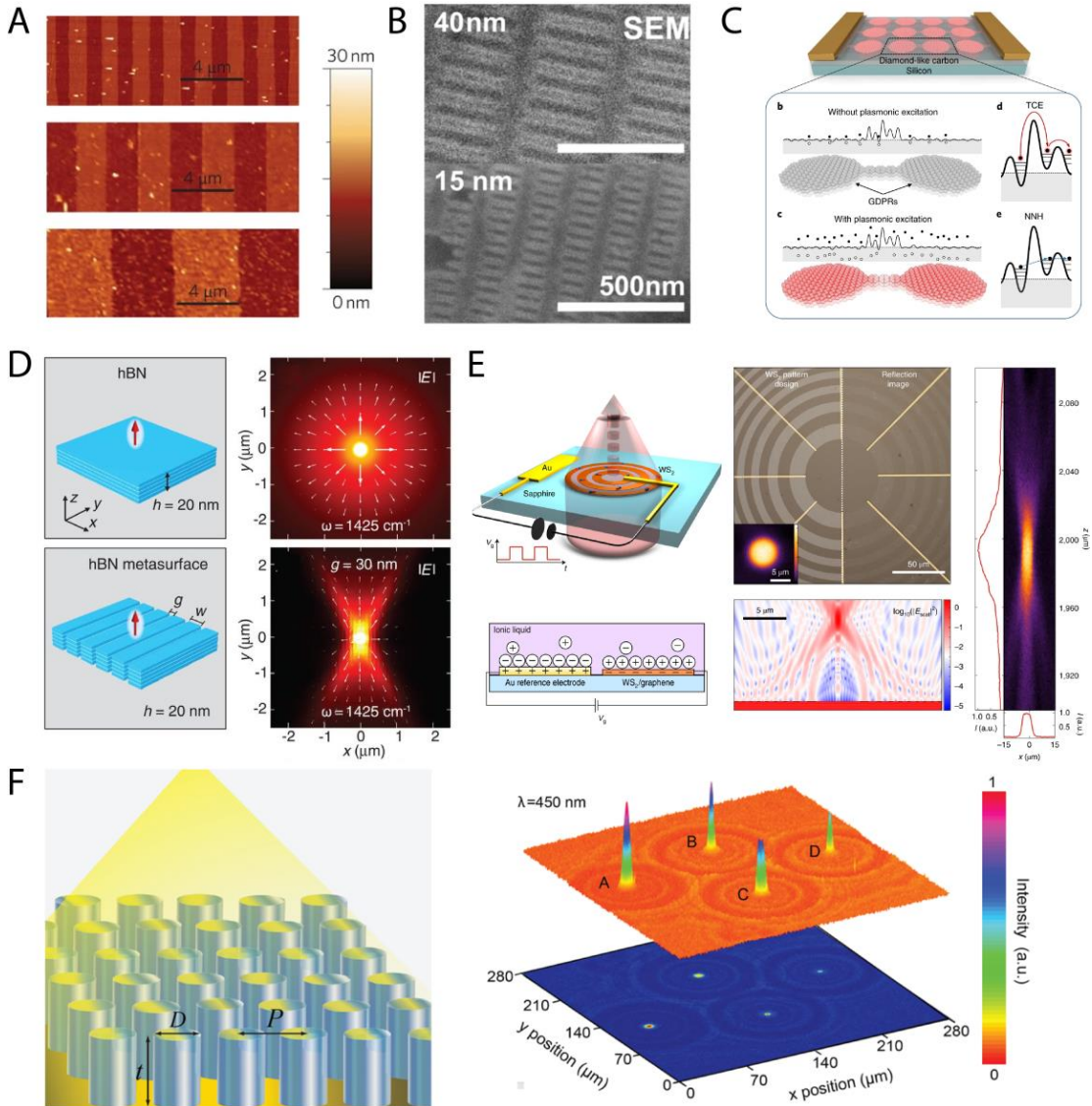

**Figure 7:** Directly patterning on 2D materials.
(A) The AFM image of a graphene micro-ribbon array, which serves as a plasmonic resonator working in THz range [253]. (B) The SEM image of a graphene tunable nano-resonator. It can be tuned by varying the back-gate voltage and width of the resonator [254]. (C) The schematic of a graphene resonator which consists of multiple graphene-disk plasmonic resonators (GDPRs) and graphene nanoribbons (GNRs) [255]. (D) The schematic of the comparison between a h-BN sheet and a patterned h-BN grating. The electric field distribution of the top surface of the h-BN excited by an out-of-plane dipole [256]. (E) The schematic of a patterned $WS_2$ monolayer that works as a metalens [239]. (F) The schematic of a h-BN flake that is patterned into a pillar array (left) can serve as a set of metalenses (right) [238].